\documentclass[reprint,prb,aps,twocolumn]{revtex4}
\usepackage[latin1]{inputenc}
\usepackage{amsmath}
\usepackage{amsfonts}
\usepackage{amssymb}
\usepackage{epsfig}
\usepackage{graphicx}
\usepackage{lscape}
\begin{document}
\title{
Accelerating Hartree--Fock exchange calculation using \\
the TURBOMOLE program system: \\ 
different techniques for different purposes \\

}
\author{Arnim Hellweg and Uwe Huniar}
\affiliation{COSMOlogic GmbH \& Co. KG,
Imbacher Weg 46,
D-51379 Leverkusen,
Germany}

\date{\today \ v1.2}

\begin{abstract}

Hartree--Fock theory is one of the most ancient methods of
computational chemistry, but up to the present day
quantum chemical calculations on Hartree--Fock level or with hybrid density 
functional theory can be excessively time consuming. 

We compare three currently available techniques to reduce the computational 
demands of such calculations in terms of timing and accuracy.

\end{abstract}
\maketitle

\section{Introduction}
\label{Intro} 

Hartree--Fock (HF) as an autonomous quantum chemical level of theory 
is nowadays hardly in use anymore, but still it 
remains one of the most applied methods, because
it serves as reference wave function for post--HF methods like
M{\o}ller--Plesset perturbation theory
(MP2, MP3, ...) or Coupled--Cluster calculations (CC2, CCSD, CCSD(T), ...).
Also, density functional theory (DFT) makes use of a HF-like method in the
computation of hybrid functionals like the popular B3-LYP or PBE0.
The evaluation of HF exchange is the time determining operation in such a 
hybrid DFT calculation. 
In modern implementations of MP2 energy calculations the HF part often 
takes significantly longer than the MP2 part.

Several techniques to reduce the computational costs 
are available on the market, e.g.
applying integral screening\cite{DSCF,LinK}, exploitation of the resolution--of--the--identity\cite{RIHF,diss:Marco} (RI), and using pseudospectral\cite{pseudo} or
seminumerical methods\cite{cosx,senex}.

There is some confusion among users about which method is
the appropriate one for their specific problem.
In this piece of writing, we summarize the efficiency and  
accuracy of available techniques and 
aim to provide some guidance for their usage.

\section{Computational background}
\label{background}

Hartree-Fock (HF) became applicable for scientific computing  
around 1950 due to the works of C.~C.~J.~Roothaan\cite{roothaan}
and G.~G.~Hall\cite{hall}  
and the introduction of spacial basis functions like those 
of Gaussian-type by Boys\cite{boys}.
The Roothaan--Hall self-consistent field method (SCF) equation is
\begin{equation}
 FC=SC\epsilon,
\end{equation}
$F$ is the Fock matrix,
$S$ is the overlap matrix,
$C$ are the expansion coefficients for the molecular orbitals (MOs),
and $\epsilon$ is a diagonal matrix with orbital energies.
The Fock matrix can be separated in the Coulomb part $J$
and the exchange part $K$. 
The SCF equation is solved iteratively, until the energy is minimized
to convergence.
$J$ and $K$ can be evaluated hereby either in shared or in separated loops
over basis functions. 
If they are calculated separately, the $J$ part can be
solved very efficiently using special techniques, e.g. with the multipole 
accelerated RI-$J$ (MARI-$J$) approach.\cite{marij}
That leaves the $K$ part to be optimized.

\subsection*{Approximations}
\label{methods}

The formal scaling of the two-electron, four center integrals which build up both 
the $J$ and $K$ part of the Fock matrix is $N^4$, with $N$ being the total 
number of atom centered basis functions (see below). 
Already direct SCF procedures with a common evaluation of $J$ and $K$ 
can be speeded up significantly and their asymptotic scaling can be reduced to $N^2$
by applying two--electron integral screening\cite{DSCF}.
The convergence can be accelerated by 
the use of minimized density differences or 
direct inversion in the iterative subspace (DIIS).\cite{diis}  
Integral screening and related techniques are generally used wherever applicable, 
since the mathematically sound deployment of upper bounds for the integrals do not 
introduce any noticeable numerical errors. While for smaller structures the formal 
scalings remains $N^4$, the situation gets better the bigger the molecules become 
- depending on the size and diffuseness of the chosen basis set (see Table \ref{tab.scaling}).

\begin{table}
\caption{Scaling behaviour of the time-demanding steps in a HF or hybrid DFT calculation}
\label{tab.scaling}
\smallskip
\begin{tabular}{ccc}\hline
  & formal scaling $\rightarrow$ & asymptotic scaling \\\hline
2-e integrals ($J$ and $K$) & $N^4$ & $N^2$ \\
( RI-$J$ instead of $J$ & $N^3$ & $N^2$ ) \\
( MARI-$J$ instead of $J$ & $N^3$ & $N$ ) \\
DFT quadrature & $N^3$ & $N$ \\
Matrix diagonalization & $N^3$ & $N^3$\\\hline
\hline \noalign{\smallskip}
\end{tabular}
\end{table}

We refer to the results from such calculations as exact HF solution,
because no approximation is used, only integrals are
neglected that do not contribute.
 
\subsubsection*{rij}
If the RI approximation is being used for the Coulomb part $J$, the operator
can be replaced: 
\begin{equation} \label{rij}
J_{\nu\mu}  \sim J_{\nu\mu}^{RI} \qquad ; \qquad K_{\nu\mu} = K_{\nu\mu}^{screened}.
\end{equation}
The $K$ part can be re-ordered for an exact optimum--screened 
procedure\cite{diss:Marco} which is similar to a linear scaling procedure as 
described by Ochsenfeld \textit{et al.}\cite{LinK}. This introduces an RI error in $J$ but not in $K$.
This approach will be denoted rij in the following.

\subsubsection*{rik}
The RI approximation can also be used to fully replace the exchange part:\cite{RIHF}
\begin{equation}
J_{\nu\mu}  \sim J_{\nu\mu}^{RI} \qquad ; \qquad K_{\nu\mu}  \sim K_{\nu\mu}^{RI}.
\end{equation}
The same Coulomb operator as above from Eq.\ref{rij} is used.
This approach will be denoted rik in the following.
(One can also call it rijk, but the $J$ part is not crucial here.)

\subsubsection*{senex}
In the approach denoted senex in the following,
again the Coulomb operator from Eq.\ref{rij} is used while the
evaluation of $K$ is now done by solving 
one integral analytically and the other one numerically on a spacial grid:\cite{senex} 
\begin{equation}
J_{\nu\mu}  \sim J_{\nu\mu}^{RI} \qquad ; \qquad K_{\nu\mu}  \sim K_{\nu\mu}^{SN}.
\end{equation}
This is a seminumerical procedure that can be done with the same 
grids on which the density functionals are evaluated.

\subsection*{Basis sets and auxiliary basis sets}
\label{basen}

The introduction of linear combination of atomic orbitals (LCAO) as
approximation of one-electron wave functions was a fundamental step
for practical calculations.
The development of general applicable basis sets made model chemistry
reproducible and revisable.
Yet, a steady and on-going development created a whole zoo of basis sets
inhibited by an enormous amount of abbreviations, lots of them only 
understandable by experts.
At the EMSL Basis Set Exchange (\verb|https://bse.pnl.gov/bse/portal|) the
most frequently used ones can be obtained ready to use in different input
formats.

Different recommendations can be given for DFT and wave function 
theories (WFT).
For DFT, already relatively small basis sets often yield decent results. 
To benchmark their performance, we use def-SV(P)\cite{Sch92} and
def-TZVP\cite{Sch94} as typical representatives.
For an improved quantitative description - especially when dealing 
with heavy element compounds -, we rather suggest to 
use the def2 basis sets\cite{Wei05}, though.
def-SV(P) is a split valence basis set with one set of polarization functions
for the non-hydrogen atoms. It is of comparable size and quality as the Pople
6-31G*\cite{Hariharan73} basis set.
def-TZVP is a triple-$\zeta$ basis set with one set of uncontracted
polarization functions. This is comparable to the 
6-311G**\cite{Krishnan80} basis set.

Wave function theory is not only more expensive in terms of computational
time, but also much more demanding concerning the basis set. 
Heavily polarized triple- and quadruple-$\zeta$ basis sets
def2-TZVPP and def2-QZVPP\cite{Wei05} as typical representatives will be used
in the following sections. Those are comparable to Dunnings cc-pVTZ and
cc-pVQZ.\cite{Dunning:89}
Additional diffuse basis functions are required when studying
anions or properties related to electron densities not localized close
to the atoms. We use def2-TZVPPD\cite{D-basen} for such a basis set.
The suffix D indicates the diffuse functions here, as the + is for the 
Pople type basis sets and the prefix aug- for the ones of Dunning type.

The prefixes def- and def2- for the basis set will be skipped in the following,
because the names are unambiguous for the elements computed in this work.

The SVP, TZVP and QZVP basis set family combines several benefits which make 
them usable for almost all applications -- namely the fact that they are
available for all elements, automatically include ECPs for heavier elements 
to include scalar relativistic effects. Furthermore, when applying the 
RI approximation, special auxiliary basis sets are needed.
For the used basis sets optimized auxiliary basis sets are available for 
RI-J, RI-K and correlated RI calculations like RI-MP2, RI-CC2, RI-CCSD, etc.
They are optimized and investigated for $J^{RI}$ in Ref.~\onlinecite{Wei06:jbasen}
and for the common usage in $J^{RI}$ and $K^{RI}$ in Ref.~\onlinecite{def2-rik}.
All TURBOMOLE basis sets and auxilary basis sets can be found online:
 \verb|http://www.cosmologic.de/basis-sets/basissets.php|

\section{Benchmark calculations}
\label{benchmark}

All calculations have been performed using TURBOMOLE V6.4.\cite{TM} 
The exact HF SCF calculations were done with the \texttt{dscf} module.
The approximated HF approaches are implemented in the \texttt{ridft}
module. (Despite of its name it can also perform non--DFT calculations.)
Default settings were used throughout if not specified otherwise.

As benchmark sets, amylose-chains containing one (24 atoms), two (45 atoms), 
and four (87 atoms) D-glucose units\cite{Kussmann07} 
and arsenic clusters As$_n$ (n=4,8,12)\cite{TM13} were selected.
$C_1$ symmetry was used in the computations.
For atomization energy tests, the structures of methane, ethane, propane,
and butane were optimized on BP/SV(P) level.

\subsection*{Accuracy}
\label{accuracy}

The differences (RI and seminumerical errors) between exact HF 
energies (using \texttt{dscf}) and the approximated HF energies (using 
\texttt{ridft}) are collected in Table \ref{tab.error} for the different 
amylose-chains with the different basis sets.

\begin{table}
\caption{The RI and seminumerical errors per atom of amylose chains in $\mu$Hartree.  
na is the number of atoms, 
nbf the number of CAO basis functions.
\label{tab.error} }
\smallskip
\begin{tabular}{lcc @{\hspace{1.2cm}} c @{\hspace{0.8cm}} c @{\hspace{0.8cm}} c} \hline
          & na     & nbf  &  rij &  rik  &  senex \\ \hline
SV(P)     &        &      &     &     &   \\
          & 24     & 204  &  146 &  -21  &  151 \\
          & 45     & 389  &  149 &  -21  &  173 \\
          & 87     & 759  &  150 &  -21  &  167 \\ \hline
TZVP      &        &      &      &     &    \\
          & 24     & 312  &  159 &  -10  &  129 \\
          & 45     & 592  &  160 &  -10  &  167 \\
          & 87     & 1152 &  160 &  -10  &  143 \\ \hline
TZVPP     &        &      &      &     &    \\
          & 24     & 612  &  31  &  -12  &  -9 \\
          & 45     & 1158 &  30  &  -12  &  23 \\
          & 87     & 2250 &  29  &  -12  &  -3 \\ \hline
TZVPPD    &        &      &      &     &    \\
          & 24     & 750  &  31  &  -12  &  -13 \\
          & 45     & 1418 &  29  &  -12  &  16  \\
          & 87     & 2754 &  29  &  -12  &  -11 \\ \hline
QZVPP     &        &      &      &     &     \\
          & 24     & 1284 &  31  &  -14  &  -51 \\
          & 45     & 2426 &  29  &  -14  &  -38 \\
          & 87     & 4710 &  29  &  -14  &  -58 \\ \hline
\hline \noalign{\smallskip}
\end{tabular}
\end{table}

rij and rik yield errors of the same size for all molecules with a specific 
basis set. This leads to an error cancellation when reactions are studied.
The errors of senex are as small as the RI errors, but they are not 
systematic for different molecular sizes.
This behavior is illustrated in Fig.~1. 
Thus, for senex no such error cancellation can be expected. 

\begin{figure}
\begin{center}
\includegraphics[width=9.9cm,angle=0]{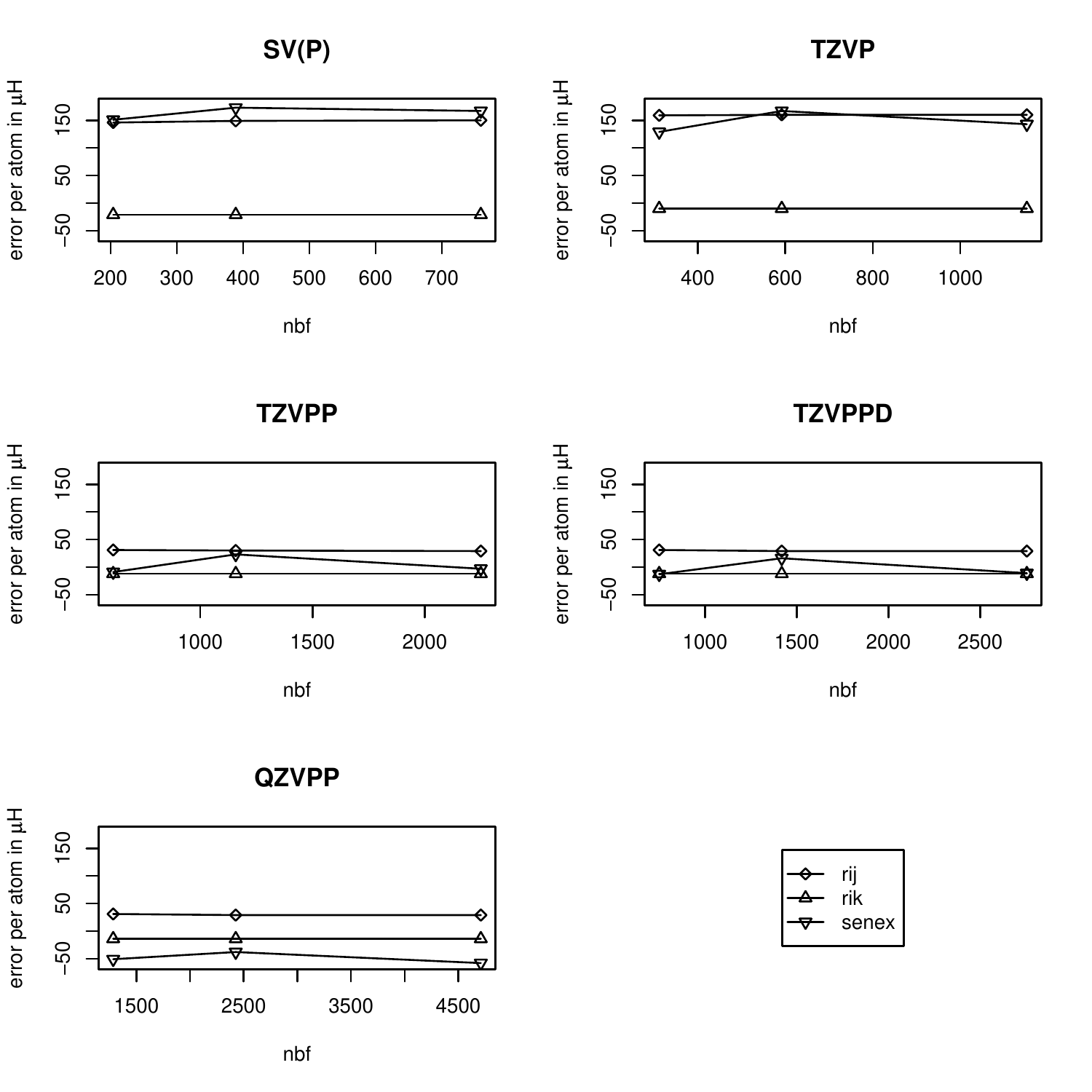}
\end{center}
\caption{The RI and seminumerical errors per atom in $\mu$Hartree for
1, 2, and 4 glucose units with the different basis sets. 
nbf is the number of CAO basis functions.
\label{fig.errors}}
\end{figure}

To get an impression of such error cancellation, atomization energies 
of small alkanes have been calculated on HF, MP2, CCSD(T), and B3-LYP level 
using the cc-pVTZ basis set. The frozen core approximation was used in
the correlated calculations. In Table \ref{tab.atomi} the differences 
(RI and seminumerical errors) between calculations using a 
reference wave function from \texttt{dscf} and from \texttt{ridft}
with the different approximations is shown.

\begin{table}
\caption{The RI and seminumerical errors in calculations of
atomization energies of alkanes using the cc-pVTZ basis set in kcal/mol. 
\label{tab.atomi} }
\smallskip
\small
\begin{tabular}{lcc@{\hspace{0.3cm}}c@{\hspace{0.3cm}}c@{\hspace{0.3cm}}c} \hline
molecule& approximation & HF   & MP2   & CCSD(T) & B3-LYP \\ \hline
methane & rij           & -0.06 & -0.06 & -0.07 & -0.07\\
        & rik           &  0.00 &  0.00 &  0.00 & 0.00\\
        & senex, m1     & -1.03 & -1.94 & -0.65 & -0.05\\
        & senex, m3     & -1.17 & -1.97 & -0.74 & -0.08\\
        & senex, m5     & -0.06 & -0.06 & -0.06 & -0.08\\ \hline
ethane  & rij           & -0.09 & -0.08 & -0.09 & -0.10\\
        & rik           &  0.00 &  0.00 &  0.00 & -0.01\\
        & senex, m1     & -2.27 & -3.92 & -1.42 & -0.10\\
        & senex, m3     & -2.30 & -3.92 & -1.45 & -0.11\\
        & senex, m5     & -0.09 & -0.08 & -0.09 & -0.11\\ \hline
propane & rij           & -0.12 & -0.11 & -0.12 & -0.13\\
        & rik           &  0.00 &  0.00 &  0.00 & -0.01\\
        & senex, m1     & -3.39 & -6.18 & -2.24 & -0.14\\
        & senex, m3     & -3.44 & -5.85 & -2.15 & -0.15\\
        & senex, m5     & -0.12 & -0.11 & -0.12 & -0.15\\ \hline
butane  & rij           & -0.15 & -0.14 & -0.15 & -0.17\\
        & rik           &  0.00 &  0.01 &  0.01 & -0.01\\
        & senex, m1     & -4.72 & -8.47 & -3.20 & -0.25\\
        & senex, m3     & -4.58 & -7.80 & -2.87 & -0.19\\
        & senex, m5     & -0.15 & -0.14 & -0.15 & -0.19\\ \hline 
\hline \noalign{\smallskip}
\end{tabular}
\end{table}

The calculation of atomization energies is known to be prone to 
insufficiencies of the basis set as well as of the method itself, since
all errors sum up. 
It can be seen that the errors introduced by the RI approximation are
well--behaved, the rik errors being an order of magnitude smaller 
than the ones from rij due to the larger auxiliary basis set of rik which 
is used for both $J$ and $K$ part of a rik calculation.
The errors of senex with the default grid (m1)
are noticeable larger and quickly increasing with the system size.
Grids of the m5 size have to be used to reach the accuracy of rij. 
This is a consequence of the irregular patterns recognizable in
Table~\ref{tab.error}.

A comparison between the different methods shows an interesting effect.
On MP2 level the errors are larger than on HF level, while they are
smaller on CCSD(T) level. It can be assumed that applying 
perturbation theory on an imprecise wave function increases its deficits, 
whereas the iterative solution of the CCSD equation repairs some of them. 

The situation looks more relaxed when using hybrid-DFT rather than pure HF.
In the last column of the table the values for the B3-LYP functional (using 
20\% HF exchange) are collected. The RI errors are of the same magnitude 
as for HF and already with small grids the seminumerical errors 
are acceptable.

\subsection*{Timings}
\label{timings}

\begin{figure*}
\begin{center}
\includegraphics[width=17.0cm,angle=0]{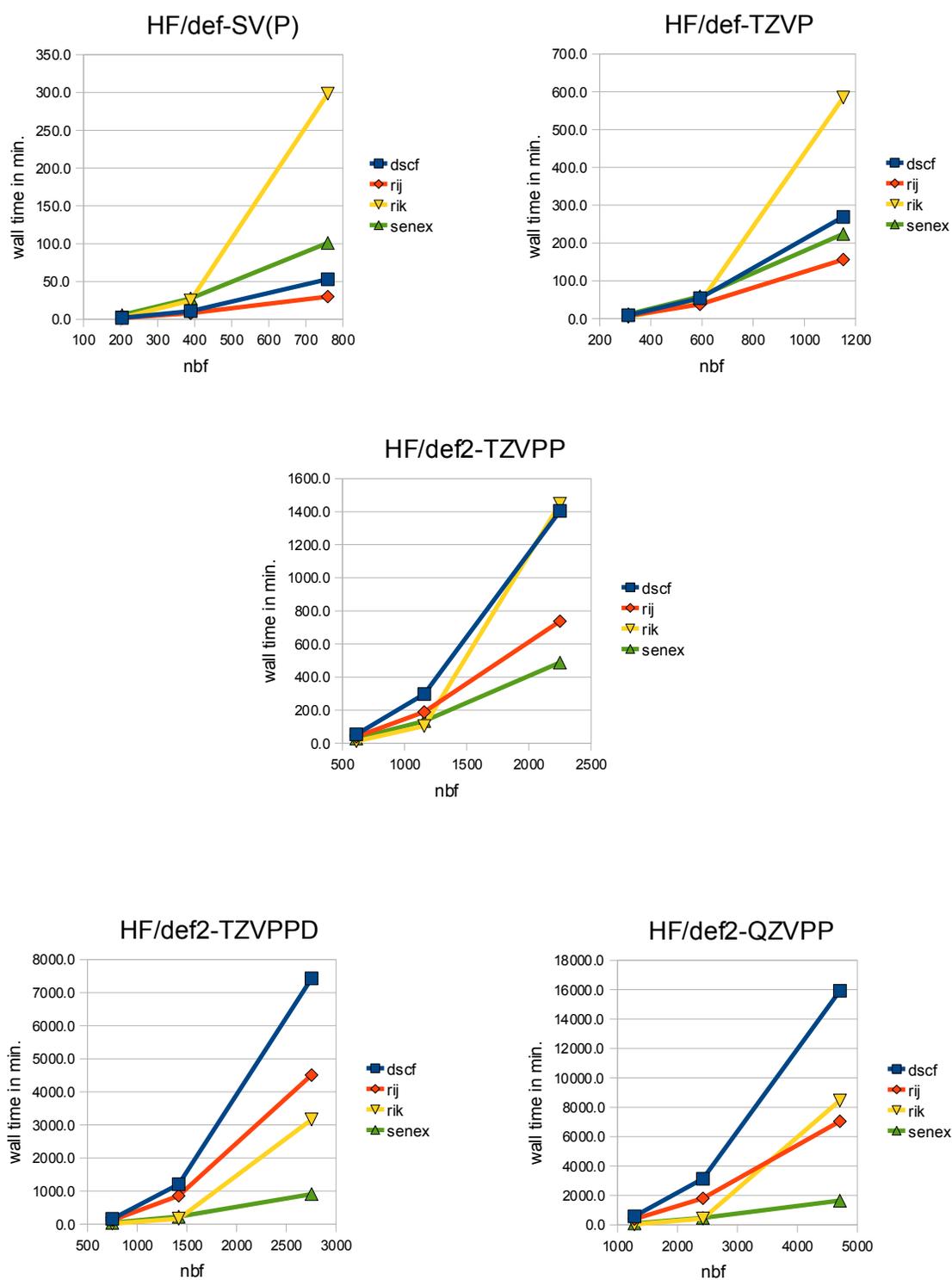}
\end{center}
\caption{The computational wall-time in minutes for
1, 2, and 4 glucose units with the different basis sets. 
nbf is the number of CAO basis functions.
\label{fig.timing}}
\end{figure*}

In Table~\ref{tab.time} the wall-times in minutes are collected
for the different approximations, basis sets, and system sizes of amylose chains. 
In Fig.~\ref{fig.timing} the data is plotted.

\begin{table}
\caption{The wall-time in minutes for HF calculations of amylose chains. 
The SCF was started from an extended H\"uckel guess.
In all cases, the number of SCF iterations in the same row was identical.
na is the number of atoms, 
nbf the number of CAO basis functions.
\label{tab.time} }
\small
\begin{tabular}{lcc@{\hspace{0.3cm}}c@{\hspace{0.1cm}}c@{\hspace{0.1cm}}c@{\hspace{0.1cm}}c@{\hspace{0.1cm}}c} \hline
          &  &   &  &  &  &  & fudged\\ 
          & na & nbf  & dscf & rij & rik & senex & rik\\ \hline
SV(P)     &        &      &      &     &     &    & \\
          & 24     & 204  & 1.9  & 1.7 & 2.3 &  5.7   &2.3\\
          & 45     & 389  & 10.8 & 8.1 & 24.9&  27.7  &24.9\\
          & 87     & 759  & 53.1 & 30.0& 298.5& 101.0 &99.5\\ \hline
TZVP      &        &      &      &     &     &     &    \\
          & 24     & 312  & 8.8  & 7.3 & 4.6 &11.8   & 4.6\\
          & 45     & 592  & 53.9 & 37.8& 42.1& 58.6  &42.1\\
          & 87     & 1152 & 269.2&156.3& 584.7& 224.0&146.2\\ \hline
TZVPP     &        &      &      &     &     &   &       \\
          &24&612&53.3&39.4&12.0&27.7 &12.0\\
          &45&1158&296.6&188.4&103.6&133.9 &103.6\\
          &87&2250&1403.9&737.6&1447.4&487.6 &180.9\\ \hline
TZVPPD    &        &      &      &     &     &  &   \\
          &24&750&159.7&121.5&16.3&40.9 &16.3\\
          &45&1418&1206.7&859.9&164.3&225.4 &82.2\\
          &87&2754&7427.1&4506.8&3162.6&908.8 &316.3\\ \hline
QZVPP     &        &      &      &     &     & &      \\
          &24&1284&582.8&388.9&33.9&98.3 &33.9\\
          &45&2426&3133.3&1808.3&427.7&470.3 &142.6\\
          &87&4710&15920.4&7043.1&8456.7&1643.3 &497.5\\ \hline
\hline \noalign{\smallskip}
\end{tabular}
\end{table}

rij always performs better than \texttt{dscf}.
For the typical DFT basis sets SV(P) and TZVP, it is the fastest
and the curve will stay under the others when going to even larger
systems due to its scaling.
For larger basis sets rik is faster for small and medium sized
systems, but rij can pass by on larger systems.

The performance of rik is very good for TZVPP basis sets or larger, 
as long as everything fits in memory. 
But performance goes down due to the need of blocking -- if the memory given
is insufficient, the algorithm has to redo the same loops for several 
batches into which the whole problem has been divided.
For the computations a maximum memory (\texttt{\$ricore}) of 2GB was chosen. 
The additional column denoted fudged rik in Table~\ref{tab.time} 
is a hypothetical timing, calculated as if enough memory were 
available such that only one batch would be sufficient.
It can be seen that the blocking is the main reason that
the performance drops for larger systems.
Since more and more memory is available in off--the--shelf computers,
rik will become more and more fetching in future applications.

senex with the m1 grid is very fast for basis sets of TZVPP size 
and larger. Especially, for bigger systems one has to admit that the 
performance looks impressive.

\subsection*{Heavier elements, truncated RI, and semi-direct \texttt{dscf}}
When studying compounds composed of elements beyond the realms of organic
chemistry, the trends and characteristics can be recognized already with smaller 
systems. We use arsenic cluster to benchmark the performance of heavier atoms.
The timings are collected in Table~\ref{tab.as-time}.
The advantage of rik when using large basis sets over dscf and rij diminishes 
for these systems compared to the amylose chains. The performance behavior
of the other approaches is quite similar to the former cases.

Two additional columns are presented here. One with timings of
semi-direct SCF as an option of dscf and the other with timings of
truncated RI as an option of rik.
In the semi-direct dscf mode, the most time consuming and frequently 
used integrals are stored on disk. 
The semi-direct mode is for smaller systems a bit faster than the fully direct
mode, but the situation changes for larger problems.

In truncated RI procedures reduced auxiliary basis sets are used 
during SCF iterations. The final SCF iteration is computed 
with the full auxiliary basis set.
Usually, functions with the two highest $l$-quantum numbers are skipped.
For rij and senex this will be not efficient, since it would only 
affect the $J$ part and it is, furthermore, not compatible with MARI-$J$.
The error of the truncation is by orders of magnitudes smaller than the RI
error on HF energies, but can increase the error of post-HF calculations,
see Ref.~\onlinecite{RIHF}.

\begin{table*}
\caption{The wall-time in minutes for HF calculations of 
arsenic cluster As$_n$ (n=4,8,12). 
The SCF was started from extended H\"uckel guess.
In all cases, the number of SCF iterations in the same row was identical.
na is the number of atoms, 
nbf the number of CAO basis functions.
\label{tab.as-time} }
\smallskip
\begin{tabular}{lcc@{\hspace{1.0cm}}c@{\hspace{0.5cm}}c@{\hspace{0.5cm}}c@{\hspace{0.5cm}}c@{\hspace{0.5cm}}c@{\hspace{0.5cm}}c@{\hspace{0.5cm}}c} \hline
&           &        &       &     &     &    & fudged & semi-direct  & trunc.\\ 
      & na    & nbf    & dscf  & rij & rik & senex & rik & dscf & rik \\ \hline 
SV(P)&&&&&&&& \\
&4&140&0.7&0.7&1.8&1.2&1.8&0.3&1.7\\
&8&280&5.5&4.7&23.5&6.9&23.5&3.1&22.0\\
&12&420&17.7&13.7&122.9&23.0&61.5&11.2&119.5\\\hline
TZVP&&&&&&&&\\
&4&156&1.5&1.4&2.1&1.6&2.1&0.5&2.5\\
&8&312&12.2&9.8&29.7&10.3&29.7&6.0&25.9\\
&12&468&34.9&25.1&135.3&30.1&67.7&20.4&121.0\\\hline
TZVPP&&&&&&&&\\
&4&220&3.8&3.3&3.3&2.4&3.3&1.8&2.9\\
&8&440&37.8&28.9&50.4&20.0&50.4&52.5&47.0\\
&12&660&105.7&75.4&212.7&53.7&70.9&106.5&186.5\\\hline
TZVPPD&&&&&&&&\\
&4&248&5.7&5.1&3.6&3.0&3.6&3.0&3.6\\
&8&496&66.4&51.0&57.7&25.3&57.7&70.7&56.1\\
&12&744&220.1&158.0&244.4&67.1&81.5&203.2&224.8\\\hline
QZVPP&&&&&&&&\\
&4&444&37.3&29.8&8.6&9.3&8.6&22.4&7.7\\
&8&888&317.0&226.1&106.6&57.9&53.3&315.9&102.5\\
&12&1332&1071.3&656.4&638.4&194.7&159.6&1093.6&588.5\\\hline
\hline \noalign{\smallskip}
\end{tabular}
\end{table*}

\section{Conclusion}
\label{conclusion}

rij is in almost all cases the better choice than plain \texttt{dscf}. 
One exception could come up for really large systems, somewhere beyond
20000 basis functions. Here, the memory demands of RI could be a 
limiting factor. However, at the moment such calculations are not yet
feasible except for single-point energy calculations.

rik is a good choice for small to medium sized molecules with large 
basis set, especially with diffuse functions.
A typical scenario is a MP2 calculation in which HF is the most time
consuming part, 
e.g. in a RI-MP2 energy calculation of glucose with def2-TZVPPD
with the \texttt{ricc2} module and a reference wave function 
from \texttt{dscf}, the HF part take 98\% of the overall run time.

From a pessimistic point of view, one could say that senex is
not fast enough for typical DFT applications with small basis sets
and not accurate enough for typical WFT applications with large basis sets.
More optimistically one has to say, that hybrid--DFT with large basis sets 
or two--component relativistic hybrid--DFT of larger system are becoming
accessible through this.

The semi-direct mode of \texttt{dscf} introduces no additionali error, but 
it is for larger systems not faster than the direct mode. Also, it has to be
noted that the semi-direct mode leads to additional I/O. 
The truncated rik is always faster than full rik, yet not really
significantly. It can be of use in hybrid--DFT calculations, but 
the small gain timing does hardly outweigh the loss of precision
to make it a good choice for WFT applications.

As general remark it should be pointed out, that
one must not mix calculations with different approximations.
In one coherent investigations all systems have to be treated in the
same way.

\section*{Acknowledgments}

We thank Florian Weigend for the courtesy of the arsenic cluster 
coordinates and critically reading the manuscript.

\bibliography{refs}

\begin{thebibliography}{24}
\expandafter\ifx\csname natexlab\endcsname\relax\def\natexlab#1{#1}\fi
\expandafter\ifx\csname bibnamefont\endcsname\relax
  \def\bibnamefont#1{#1}\fi
\expandafter\ifx\csname bibfnamefont\endcsname\relax
  \def\bibfnamefont#1{#1}\fi
\expandafter\ifx\csname citenamefont\endcsname\relax
  \def\citenamefont#1{#1}\fi
\expandafter\ifx\csname url\endcsname\relax
  \def\url#1{\texttt{#1}}\fi
\expandafter\ifx\csname urlprefix\endcsname\relax\def\urlprefix{URL }\fi
\providecommand{\bibinfo}[2]{#2}
\providecommand{\eprint}[2][]{\url{#2}}

\bibitem[{\citenamefont{H{\"a}ser and Ahlrichs}(1989)}]{DSCF}
\bibinfo{author}{\bibfnamefont{M.}~\bibnamefont{H{\"a}ser}} \bibnamefont{and}
  \bibinfo{author}{\bibfnamefont{R.}~\bibnamefont{Ahlrichs}},
  \bibinfo{journal}{J. Comput. Chem.} \textbf{\bibinfo{volume}{10}},
  \bibinfo{pages}{104} (\bibinfo{year}{1989}).

\bibitem[{\citenamefont{Ochsenfeld et~al.}(1998)\citenamefont{Ochsenfeld,
  White, and Head-Gordon}}]{LinK}
\bibinfo{author}{\bibfnamefont{C.}~\bibnamefont{Ochsenfeld}},
  \bibinfo{author}{\bibfnamefont{C.~A.} \bibnamefont{White}}, \bibnamefont{and}
  \bibinfo{author}{\bibfnamefont{M.}~\bibnamefont{Head-Gordon}},
  \bibinfo{journal}{J. Chem. Phys.} \textbf{\bibinfo{volume}{109}},
  \bibinfo{pages}{1663} (\bibinfo{year}{1998}).

\bibitem[{\citenamefont{Weigend}(2002)}]{RIHF}
\bibinfo{author}{\bibfnamefont{F.}~\bibnamefont{Weigend}},
  \bibinfo{journal}{Phys. Chem. Chem. Phys.} \textbf{\bibinfo{volume}{4}},
  \bibinfo{pages}{4285} (\bibinfo{year}{2002}).

\bibitem[{\citenamefont{Kattannek}(2006)}]{diss:Marco}
\bibinfo{author}{\bibfnamefont{M.}~\bibnamefont{Kattannek}}, Ph.D. thesis,
  \bibinfo{school}{Fakult\"at f\"ur Chemie und Biowissenschaften, Universit\"at
  Karlsruhe (TH)} (\bibinfo{year}{2006}).

\bibitem[{\citenamefont{Friesner}(1985)}]{pseudo}
\bibinfo{author}{\bibfnamefont{R.~A.} \bibnamefont{Friesner}},
  \bibinfo{journal}{Chem. Phys. Lett.} \textbf{\bibinfo{volume}{116}},
  \bibinfo{pages}{39} (\bibinfo{year}{1985}).

\bibitem[{\citenamefont{Neese et~al.}(2009)\citenamefont{Neese, Wennmohs,
  Hansen, and Becker}}]{cosx}
\bibinfo{author}{\bibfnamefont{F.}~\bibnamefont{Neese}},
  \bibinfo{author}{\bibfnamefont{F.}~\bibnamefont{Wennmohs}},
  \bibinfo{author}{\bibfnamefont{A.}~\bibnamefont{Hansen}}, \bibnamefont{and}
  \bibinfo{author}{\bibfnamefont{U.}~\bibnamefont{Becker}},
  \bibinfo{journal}{Chem. Phys.} \textbf{\bibinfo{volume}{356}},
  \bibinfo{pages}{98} (\bibinfo{year}{2009}).

\bibitem[{\citenamefont{Plessow and Weigend}(2012)}]{senex}
\bibinfo{author}{\bibfnamefont{P.}~\bibnamefont{Plessow}} \bibnamefont{and}
  \bibinfo{author}{\bibfnamefont{F.}~\bibnamefont{Weigend}},
  \bibinfo{journal}{J. Comput. Chem.} \textbf{\bibinfo{volume}{33}},
  \bibinfo{pages}{810} (\bibinfo{year}{2012}).

\bibitem[{\citenamefont{Roothaan}(1951)}]{roothaan}
\bibinfo{author}{\bibfnamefont{C.~C.~J.} \bibnamefont{Roothaan}},
  \bibinfo{journal}{Rev. Mod. Phys.} \textbf{\bibinfo{volume}{23}},
  \bibinfo{pages}{69} (\bibinfo{year}{1951}).

\bibitem[{\citenamefont{Hall}(1951)}]{hall}
\bibinfo{author}{\bibfnamefont{G.~G.} \bibnamefont{Hall}},
  \bibinfo{journal}{Proc. R. Soc. Lond. A} \textbf{\bibinfo{volume}{205}},
  \bibinfo{pages}{541} (\bibinfo{year}{1951}).

\bibitem[{\citenamefont{Boys}(1950)}]{boys}
\bibinfo{author}{\bibfnamefont{S.~F.} \bibnamefont{Boys}},
  \bibinfo{journal}{Proc. R. Soc. Lond. A} \textbf{\bibinfo{volume}{200}},
  \bibinfo{pages}{542} (\bibinfo{year}{1950}).

\bibitem[{\citenamefont{Sierka et~al.}(2003)\citenamefont{Sierka, Hogekamp, and
  Ahlrichs}}]{marij}
\bibinfo{author}{\bibfnamefont{M.}~\bibnamefont{Sierka}},
  \bibinfo{author}{\bibfnamefont{A.}~\bibnamefont{Hogekamp}}, \bibnamefont{and}
  \bibinfo{author}{\bibfnamefont{R.}~\bibnamefont{Ahlrichs}},
  \bibinfo{journal}{J. Chem. Phys.} \textbf{\bibinfo{volume}{118}},
  \bibinfo{pages}{9136} (\bibinfo{year}{2003}).

\bibitem[{\citenamefont{Pulay}(1980)}]{diis}
\bibinfo{author}{\bibfnamefont{P.}~\bibnamefont{Pulay}},
  \bibinfo{journal}{Chem. Phys. Lett.} \textbf{\bibinfo{volume}{73}},
  \bibinfo{pages}{393} (\bibinfo{year}{1980}).

\bibitem[{\citenamefont{Sch\"afer et~al.}(1992)\citenamefont{Sch\"afer, Horn,
  and Ahlrichs}}]{Sch92}
\bibinfo{author}{\bibfnamefont{A.}~\bibnamefont{Sch\"afer}},
  \bibinfo{author}{\bibfnamefont{H.}~\bibnamefont{Horn}}, \bibnamefont{and}
  \bibinfo{author}{\bibfnamefont{R.}~\bibnamefont{Ahlrichs}},
  \bibinfo{journal}{J. Chem. Phys.} \textbf{\bibinfo{volume}{97}},
  \bibinfo{pages}{2571} (\bibinfo{year}{1992}).

\bibitem[{\citenamefont{Sch\"afer et~al.}(1994)\citenamefont{Sch\"afer, Huber,
  and Ahlrichs}}]{Sch94}
\bibinfo{author}{\bibfnamefont{A.}~\bibnamefont{Sch\"afer}},
  \bibinfo{author}{\bibfnamefont{C.}~\bibnamefont{Huber}}, \bibnamefont{and}
  \bibinfo{author}{\bibfnamefont{R.}~\bibnamefont{Ahlrichs}},
  \bibinfo{journal}{J. Chem. Phys.} \textbf{\bibinfo{volume}{100}},
  \bibinfo{pages}{5829} (\bibinfo{year}{1994}).

\bibitem[{\citenamefont{Weigend and Ahlrichs}(2005)}]{Wei05}
\bibinfo{author}{\bibfnamefont{F.}~\bibnamefont{Weigend}} \bibnamefont{and}
  \bibinfo{author}{\bibfnamefont{R.}~\bibnamefont{Ahlrichs}},
  \bibinfo{journal}{Phys. Chem. Chem. Phys.} \textbf{\bibinfo{volume}{7}},
  \bibinfo{pages}{3297} (\bibinfo{year}{2005}).

\bibitem[{\citenamefont{Hariharan and Pople}(1973)}]{Hariharan73}
\bibinfo{author}{\bibfnamefont{P.~C.} \bibnamefont{Hariharan}}
  \bibnamefont{and} \bibinfo{author}{\bibfnamefont{J.~A.} \bibnamefont{Pople}},
  \bibinfo{journal}{Theor. Chem. Acc.} \textbf{\bibinfo{volume}{28}},
  \bibinfo{pages}{213} (\bibinfo{year}{1973}).

\bibitem[{\citenamefont{Krishnan et~al.}(1980)\citenamefont{Krishnan, Binkley,
  Seeger, and Pople}}]{Krishnan80}
\bibinfo{author}{\bibfnamefont{R.}~\bibnamefont{Krishnan}},
  \bibinfo{author}{\bibfnamefont{J.~S.} \bibnamefont{Binkley}},
  \bibinfo{author}{\bibfnamefont{R.}~\bibnamefont{Seeger}}, \bibnamefont{and}
  \bibinfo{author}{\bibfnamefont{J.~A.} \bibnamefont{Pople}},
  \bibinfo{journal}{J. Chem. Phys.} \textbf{\bibinfo{volume}{72}},
  \bibinfo{pages}{650} (\bibinfo{year}{1980}).

\bibitem[{\citenamefont{Dunning}(1989)}]{Dunning:89}
\bibinfo{author}{\bibfnamefont{T.~H.} \bibnamefont{Dunning}},
  \bibinfo{journal}{J. Chem. Phys.} \textbf{\bibinfo{volume}{90}},
  \bibinfo{pages}{1007} (\bibinfo{year}{1989}).

\bibitem[{\citenamefont{Rappoport and Furche}(2010)}]{D-basen}
\bibinfo{author}{\bibfnamefont{D.}~\bibnamefont{Rappoport}} \bibnamefont{and}
  \bibinfo{author}{\bibfnamefont{F.}~\bibnamefont{Furche}},
  \bibinfo{journal}{J. Chem. Phys.} \textbf{\bibinfo{volume}{133}},
  \bibinfo{pages}{134105} (\bibinfo{year}{2010}).

\bibitem[{\citenamefont{Weigend}(2006)}]{Wei06:jbasen}
\bibinfo{author}{\bibfnamefont{F.}~\bibnamefont{Weigend}},
  \bibinfo{journal}{Phys. Chem. Chem. Phys.} \textbf{\bibinfo{volume}{8}},
  \bibinfo{pages}{1057} (\bibinfo{year}{2006}).

\bibitem[{\citenamefont{Weigend}(2008)}]{def2-rik}
\bibinfo{author}{\bibfnamefont{F.}~\bibnamefont{Weigend}}, \bibinfo{journal}{J.
  Comput. Chem.} \textbf{\bibinfo{volume}{29}}, \bibinfo{pages}{167}
  (\bibinfo{year}{2008}).

\bibitem[{TM()}]{TM}
\emph{\bibinfo{title}{{TURBOMOLE V6.4 2012}, a development of {University of
  Karlsruhe} and {Forschungszentrum Karlsruhe GmbH}, 1989-2007, {TURBOMOLE
  GmbH}, since 2007; available from \\ {\tt http://www.turbomole.com}.}}

\bibitem[{\citenamefont{Kussmann and Ochsenfeld}(2007)}]{Kussmann07}
\bibinfo{author}{\bibfnamefont{J.}~\bibnamefont{Kussmann}} \bibnamefont{and}
  \bibinfo{author}{\bibfnamefont{C.}~\bibnamefont{Ochsenfeld}},
  \bibinfo{journal}{J. Chem. Phys.} \textbf{\bibinfo{volume}{127}},
  \bibinfo{pages}{054103} (\bibinfo{year}{2007}).

\bibitem[{\citenamefont{Furche et~al.}(2014)\citenamefont{Furche, Ahlrichs,
  H{\"a}ttig, Klopper, Sierka, and Weigend}}]{TM13}
\bibinfo{author}{\bibfnamefont{F.}~\bibnamefont{Furche}},
  \bibinfo{author}{\bibfnamefont{R.}~\bibnamefont{Ahlrichs}},
  \bibinfo{author}{\bibfnamefont{C.}~\bibnamefont{H{\"a}ttig}},
  \bibinfo{author}{\bibfnamefont{W.}~\bibnamefont{Klopper}},
  \bibinfo{author}{\bibfnamefont{M.}~\bibnamefont{Sierka}}, \bibnamefont{and}
  \bibinfo{author}{\bibfnamefont{F.}~\bibnamefont{Weigend}},
  \bibinfo{journal}{WIREs Comput Mol Sci} \textbf{\bibinfo{volume}{4}},
  \bibinfo{pages}{91} (\bibinfo{year}{2014}).

\end{thebibliography}

\end{document}